\documentclass[twocolumn,tighten]{aastex63}

\usepackage{graphicx}
\usepackage{ulem}
\usepackage{amsmath}
\usepackage{wrapfig}
\usepackage[thinlines]{easytable}
\usepackage{tikz}

\newcolumntype{P}[1]{>{\centering\arraybackslash}p{#1}}

\newcommand\lsim{\mathrel{\rlap{\lower4pt\hbox{\hskip1pt$\sim$}}
        \raise1pt\hbox{$<$}}}
\newcommand\gsim{\mathrel{\rlap{\lower4pt\hbox{\hskip1pt$\sim$}}
        \raise1pt\hbox{$>$}}}

\begin{document}
\shorttitle{TDEs in young star clusters}
\shortauthors{}

\title{Fast Optical Transients from Stellar-Mass Black Hole Tidal Disruption Events in Young Star Clusters}
%% Authors with the same affiliation can be grouped in a single
%% \author and \affil call.

\correspondingauthor{Kyle Kremer}
\email{kkremer@caltech.edu}

\author[0000-0002-4086-3180]{Kyle Kremer}
\altaffiliation{NSF Astronomy \& Astrophysics Postdoctoral Fellow}
\affiliation{TAPIR, California Institute of Technology, Pasadena, CA 91125, USA}
\affiliation{The Observatories of the Carnegie Institution for Science, Pasadena, CA 91101, USA}

\author[0000-0002-1568-7461]{Wenbin Lu}
\affiliation{TAPIR, California Institute of Technology, Pasadena, CA 91125, USA}

\author[0000-0001-6806-0673]{Anthony L. Piro}
\affiliation{The Observatories of the Carnegie Institution for Science, Pasadena, CA 91101, USA}

\author[0000-0002-3680-2684]{Sourav Chatterjee}
\affil{Tata Institute of Fundamental Research, Homi Bhabha Road, Navy Nagar, Colaba, Mumbai 400005, India}

\author[0000-0002-7132-418X]{Frederic A. Rasio}
\affil{Center for Interdisciplinary Exploration \& Research in Astrophysics (CIERA) and Department of Physics \& Astronomy, Northwestern University, Evanston, IL 60208, USA}

\author[0000-0001-9582-881X]{Claire S. Ye}
\affil{Center for Interdisciplinary Exploration \& Research in Astrophysics (CIERA) and Department of Physics \& Astronomy, Northwestern University, Evanston, IL 60208, USA}

\begin{abstract}
Observational evidence suggests that the majority of stars may have been born in stellar clusters or associations. Within these dense environments, dynamical interactions lead to high rates of close stellar encounters. A variety of recent observational and theoretical indications suggest stellar-mass black holes may be present and play an active dynamical role in stellar clusters of all masses. In this study, we explore the tidal disruption of main sequence stars by stellar-mass black holes in young star clusters. We compute a suite of over 3000 independent $N$-body simulations that cover a range in cluster mass, metallicity, and half-mass radii. We find stellar-mass black hole tidal disruption events (TDEs) occur at an overall rate of up to roughly $200\,\rm{Gpc}^{-3}\,\rm{yr}^{-1}$ in young stellar clusters in the local universe. These TDEs are expected to have several characteristic features, namely fast rise times of order a day, peak X-ray luminosities of at least $10^{44}\,\rm{erg\,s}^{-1}$, and bright optical luminosities (roughly $10^{41}-10^{44}\,\rm{erg\,s}^{-1}$) associated with reprocessing by a disk wind. In particular, we show these events share many features in common with the emerging class of Fast Blue Optical Transients. 
\vspace{1cm}
\end{abstract}

\section{Introduction}
\label{sec:intro}

The majority of stars are expected to form in clustered environments such as young star clusters \citep[YSCs; e.g.,][]{Carpenter2000,LadaLada2003}. Several examples of YSCs exist in the Milky Way and the Local Group, and they are expected to be particularly abundant in starburst and interacting galaxies \citep[e.g.,][]{PortegiesZwart2010}. As dense stellar systems, YSCs undergo intense dynamical evolution governed by two-body relaxation, similar to their globular cluster (GC) cousins. Unlike GCs which are massive ($\sim 10^5-10^6\,M_{\odot}$) and old (ages of $10\,$Gyr or more), YSCs are generally low mass ($\lesssim 10^5\,M_{\odot}$) and short lived -- many dissolve in the disk of their host galaxy on time scales of $\mathcal{O}(100\,\rm{Myr})$ \citep[e.g.,][]{Kruijssen2011}. However, before they dissolve, the stellar dynamical processes operating in YSCs make them efficient nurseries for many unusual astrophysical objects.

Over the past decade, the topic of stellar-mass black hole (BH) populations in stellar clusters has seen a boom in interest. On the observational side, a growing number of stellar-mass BH candidates have been observed in Milky Way GCs through both radial velocity measurements \citep{Giesers2018,Giesers2019} and through X-ray/radio observations \citep{Maccarone2007,Strader2012,Chomiuk2013,Miller-Jones2015,Shishkovsky2018}. On the theoretical side, state-of-the-art $N$-body modelling has shown that stellar-mass BHs form and are retained in stellar clusters of all masses \citep[e.g.,][]{Morscher2015,Wang2015,Rodriguez2016b,Banerjee2017,Askar2018,ArcaSedda2018,Kremer2020,Weatherford2019}. Furthermore, cluster simulations have revealed that these BHs play a crucial role in the long-term dynamics, core evolution, and survival of stellar clusters \citep[e.g.,][]{Mackey2007,BreenHeggie2013,Chatterjee2017a,Kremer2018b,Ye2018,Kremer2019a, Giersz2019,Kremer2020,Wang2020}.

One of the most exciting developments in this field lies in gravitational wave (GW) astrophysics. After formation, BHs are expected to rapidly sink to the center of their host cluster through dynamical friction \citep[e.g.,][]{Kulkarni1993,Sigurdsson1993,Morscher2015}. Within their host cluster's dense core, BH--BH binaries form and subsequently harden through three-body dynamical encounters. Ultimately (depending on the host cluster's escape velocity), these binary BHs (BBHs) are either dynamically ejected from their host cluster through gravitational recoil or merge inside their host cluster through GW inspiral.
Recent studies have shown that YSCs \citep[e.g.,][]{Ziosi2014,DiCarlo2019,Banerjee2020} and old GCs \citep[e.g.,][]{Rodriguez2016a,RodriguezLoeb2018,Kremer2020,Antonini2019} may contribute comparably to the overall BBH merger rate.

Additionally, stellar-mass BHs are expected to dynamically interact with luminous stars in stellar clusters. BH--star encounters are expected to play a crucial role in the formation of both accreting and detached BH binaries \citep[e.g.,][]{Ivanova2010,Ivanova2017,Kremer2018a,Giesler2018} with properties similar to the BH-candidates detected to date in Milky Way GCs \citep[e.g.,][]{Kremer2019a}. Additionally, such dynamical encounters may occasionally cause a star to cross a BH within its tidal disruption radius, leading to a tidal disruption of the star \citep{Perets2016,Kremer2019c,Lopez2018,Samsing2019,Kremer2019e,Fragione2020}.
These stellar-mass BH tidal disruption events (TDEs) may occur during close encounters of pairs of single stars (i.e., single--single interactions) and also during small $N$ (typically three- or four-body) resonant encounters that occur through binary-mediated dynamical interactions \citep[e.g.,][]{Fregeau2007}.

Regardless of the dynamical pathway, these TDEs are expected to lead to transients with fast rise times of roughly a day driven by viscous accretion onto the BH \citep[e.g.,][]{Perets2016}. Depending on the assumed accretion efficiency of the subsequently formed accretion disk, outflows associated with disk wind mass loss are expected to re-process the inner-disk radiation on a timescale of a few to 10 days, leading to peak bolometric luminosities up to roughly $10^{44}\,\rm{erg\,s}^{-1}$ that peak in the optical \citep{Kremer2019c}. In the case of extremely efficient energy release in the form of a jet, a bright X-ray or $\gamma$-ray flare may result with the overall phenomenology possibly resembling ultra-long gamma ray bursts \citep{Perets2016}. These TDEs are expected to occur in GCs at rates of roughly $3-10\,\rm{Gpc}^{-3}\,\rm{yr}^{-1}$ \citep{Perets2016,Lopez2018,Kremer2019c} and at similar rates in both nuclear star clusters \citep{Fragione2020} and in stellar triples under the influence of Lidov-Kozai oscillations \citep{Fragione2019}.

In this paper, we examine the tidal disruption of main sequence stars by stellar-mass BHs in YSCs, in particular investigating the low-mass cluster regime, which is expected to dominate (by total number of clusters) the overall cluster mass function. This expands upon earlier work on the topic which was limited to old and massive GCs. We compile an extensive suite of $N$-body cluster models for various cluster masses and explore both TDE rates as well as characteristic properties.

Recent, current, and upcoming high-cadence surveys such as the Palomar Transient Factory \citep[e.g.,][]{PTF2009}, the Zwicky Transient Facility \citep[e.g.,][]{ZTF2019}, ASAS-SN \citep[e.g.,][]{Shappee2014}, ATLAS \citep[e.g.,][]{ATLAS2018}, Pan-STARRs \citep[e.g.,][]{PanSTARRs2016}, and the Vera Rubin Observatory \citep[e.g.,][]{LSST2009} are ushering in an unprecedented era in transient astronomy. Thus, the catalog of observed transients of both known and unknown origin is growing and will continue to grow rapidly. We conclude this study by examining the electromagentic features of stellar-mass BH TDEs and compare these events specifically with the emerging class of Fast Blue Optical Transients \citep[FBOTs; e.g.,][]{Drout2014,Arcavi2016,Rest2018,Pursiainen2018,Margutti2019,Ho2020,Coppejans2020}. On the basis of event rates, host galaxy properties, and overall transient features such as rise times and peak luminosities, we demonstrate stellar-mass BH TDEs may indeed be a viable mechanism for FBOT-like events.

In Section \ref{sec:methods}, we describe the methods we use to model stellar clusters. In Sections \ref{sec:single_BH_TDEs} and \ref{sec:BBH_TDEs}, we estimate TDE rates occurring through single--single and binary-mediated encounters, respectively, and compare the rates estimated from our $N$-body models with simple analytic estimates. In Section \ref{sec:rates}, we compute the overall TDE rates at various cosmological distances and in Section \ref{sec:masses}, we describe how TDE properties may vary with cluster metallicity. In Section \ref{sec:EMsignatures}, we discuss of the expected outcome of stellar-mass BH TDEs, specifically describing the basic properties of disk formation and evolution and the associated electromagnetic signatures. We also compare these features with observed properties of FBOTs. We discuss our results and conclude in Section \ref{sec:conclusion}.

 \section{N-body models of young clusters}
\label{sec:methods}

\begin{deluxetable*}{l|c|c|c|c|c|c|c|c}
\tabletypesize{\footnotesize}
\tablewidth{0pt}
\tablecaption{List of $N$-body models \label{table:models}}
\tablehead{
    \colhead{$^1$Label} &
	\colhead{$^2 N$} &
	\colhead{$^3 M_{\rm{cl}}$} &
	\colhead{$^4$num. of models} &
	\colhead{$^5 r_v$}&
	\colhead{$^6 Z$}&
	\colhead{$^7 t_{\rm{max}}$} &
	\colhead{$^8$single--single TDEs} &
	\colhead{$^9$binary-mediated TDEs}\\
	\colhead{} &
	\colhead{($\times10^4)$} &
	\colhead{($\times10^4\,M_{\odot})$} &
	\colhead{} &
	\colhead{(pc)} &
	\colhead{($Z_{\odot}$)} &
	\colhead{(Myr)} &
	\colhead{} &
	\colhead{}
}
\startdata
\texttt{a} & 1 & 0.6 & 1000 & 1 & 1 & 150 & 6 & 11\\
\texttt{b} & 2 & 1.2 & 400 & 1 & 1 & 150 & 8 & 7\\
\texttt{c} & 4 & 2.4 & 300 & 1 & 1 & 150 & 11 & 9\\
\texttt{d} & 6 & 3.6 & 300 & 1 & 1 & 500 & 75 & 91\\
\texttt{e} & 8 & 4.8 & 200 & 1 & 1 & 500 & 83 & 59\\
\texttt{f} & 10 & 6.0 & 150 & 1 & 1 & 500 & 100 & 53\\
\texttt{g} & 20 & 12 & 100 & 1 & 1 & 500 & 187 & 61\\
\texttt{h} & 80 & 48 & 10 & 1 & 1 & 500 & 182 & 15\\
\hline
\texttt{i} & 8 & 4.8 & 200 & 1 & 0.1 & 500 & 122 & 78\\
\texttt{j} & 10 & 6.0 & 150 & 1 & 0.1 & 500 & 170 & 58\\
\texttt{k} & 20 & 12 & 100 & 1 & 0.1 & 500 & 238 & 61\\
\hline
\texttt{l} & 10 & 6.0 & 50 & 0.42 & 1 & 500 & 128 & 138\\
\texttt{m} & 10 & 6.0 & 50 & 0.42 & 0.1 & 500 & 156 & 61\\
\hline
\texttt{n} & 10 & 6.0 & 50 & 2 & 1 & 500 & 5 & 3 \\
\hline
\hline
\texttt{o}$^{\star}$ & 12.5 & 6.0 & 50 & 1 & 1 & 500 & 0 & 229\\
\enddata
\tablecomments{All models computed in this study. In columns 2 and 3 we list the initial number of stars and cluster mass, respectively. In column 4, we list the total number of independent realizations computed for the given set of initial conditions. In columns 5, 6, and 7, we list the initial virial radius, metallicity, and maximum integration time, respectively. In columns 8 and 9, we list the total number of TDEs occurring through single--single and binary--single encounters respectively. $^\star$Unlike models \texttt{a-n} which adopt zero primordial binaries, model \texttt{o} adopts a $100\%$ primordial binary fraction. We discuss this further in Section \ref{sec:conclusion}.}
\end{deluxetable*}

To model the evolution of YSCs, we use the H\'{e}non-type Monte Carlo code \texttt{CMC} \citep{Joshi2000,Pattabiraman2013,Kremer2020}. \texttt{CMC} includes various physical processes necessary to study both large scale cluster dynamics and the formation and evolution of stellar-mass BHs, including two-body relaxation, stellar and binary star evolution \citep[computed using updated versions of \texttt{SSE} and \texttt{BSE};][]{Hurley2000,Hurley2002}, and direct integration of small-$N$ resonate encounters \citep{Fregeau2007} including post-Newtonian effects \citep{Rodriguez2018}.

To compute compact object (BH and neutron star; NS) masses, we adopt the stellar wind prescriptions of \citet{Vink2001} to determine the final stellar mass at the moment of core collapse and adopt the ``Rapid'' supernova explosion models \citep{Fryer2012} to compute NS and BH masses modified to include the prescriptions for (pulsational) pair-instability supernovae described in \citet{Belczynski2016b}. BH and NS natal kicks are computed as in \citet{Kremer2020}.

In order to treat stellar-mass BH TDEs, we adopt the same prescriptions as in \citet{Kremer2019c}. In short, if a dynamical encounter involving at least one BH and one star\footnote{Here, we are specifically interested in disruption of \textit{main sequence} stars and do not consider the disruption of giants which occur roughly a factor of 10 less frequently \citep[e.g.,][]{Kremer2019c}. Henceforth, we use the term ``star'' to mean a main sequence star. For discussion of the interaction of BHs with giants, see \citet{Ivanova2010,Ivanova2017,Kremer2019c}.} leads to a BH--star pericenter passage, $r_p$, within the star's tidal disruption radius

\begin{equation}
\label{eq:r_TD}
r_{\rm{TD}} = \Big( \frac{m_{\rm{BH}}}{m_{\star}}\Big)^{1/3}\,R_{\star},
\end{equation}
where $m_{\rm{BH}}$ is the BH mass, and $m_{\star}$ and $R_{\star}$ are the stellar mass and radius, respectively, we assume a TDE occurs.\footnote{In reality, the tidal disruption radius of a particular object likely depends also upon the object's stellar structure. In particular, this dependence may change as stars evolve and develop a more pronounced core-envelope structure. We reserve inclusion of these more detailed effects for future study and note that these effects are unlikely to affect the results presented here significantly.} At this point, we record the stellar properties and then assume the star is instantaneously destroyed.
In reality, especially if the TDE occurs during a multi-body resonant encounter, TDEs may affect the hydrodynamic evolution of their dynamical encounters. These more complex effects are well beyond the computational scope of an $N$-body code like \texttt{CMC}, but see e.g., \citet{Lopez2018} for discussion.

We apply this TDE prescription only for BH--star interactions. For close encounters of star--star pairs, we allow the stars to interact only in the direct collision limit: we assume a sticky sphere collision (i.e., zero mass loss) occurs if $r_p <R_1+R_2$, where $R_1$ and $R_2$ are the stellar radii. See \citet{Kremer2020c} for further details of our treatment of star--star collisions. We record TDEs/collisions that occur during both single--single encounters and binary-mediated dynamical encounters that are integrated directly using \texttt{Fewbody}. For further detail, see \citet{Fregeau2007, Kremer2019c,Kremer2020c}.

In all models, we assume a static Milky-Way-like external tidal field representative of the solar neighborhood (i.e., located at a distance of $8\,$kpc from the Galactic center). In reality, this choice likely underestimates the role of external tides on the long-term cluster evolution as it does not incorporate the effects of massive perturbers (e.g. molecular clouds), which may accelerate the cluster disruption \citep[e.g.,][]{Gieles2006}. We do not model here the dynamics of the final phase of cluster dissolution. As in \citet{DiCarlo2019}, we integrate our clusters to a maximum age of $150-500\,$Myr, depending on the cluster mass \citep[allowing more massive clusters to evolve longer to reflect their long relaxation times; e.g.,][]{HeggieHut2003}.
Indeed, assuming a maximum age of $500\,$Myr is appropriate as our focus here lies primarily on young clusters as opposed to long-lived globular clusters with ages of $10\,$Gyr or more \citep[see][for further discussion]{PortegiesZwart2010}. Furthermore, real star clusters are formed in a complicated interaction between gas and gravity \citep[e.g.,][]{Bate2003}, which, in general, is poorly understood. In \texttt{CMC}, we neglect the initial gas-rich phase of cluster evolution and instead assume a single starburst creates all stars. In particular, we do not consider expulsion of primordial gas that occurs on a dynamical timescale at early times and the possible consequences on the cluster dynamics/survival \citep[e.g., the ``infant mortality'' effect;][]{LadaLada2003}.

We consider initial cluster masses in the range $6000-4.8\times10^5\,M_{\odot}$, reflective of the YSC masses observed in local universe \citep[e.g.,][]{LadaLada2003,PortegiesZwart2010}. For all models, we adopt a standard \citet{Kroupa2001} initial mass function with a mass range of $0.08-150\,M_{\odot}$. We assume all models are initially fit to a King model with concentration parameter $W_0=5$ \citep{King1962}. 

We adopt two values for the cluster initial virial radius, $r_v$. In the first limit, we assume a constant $r_v=1\,$pc for all cluster masses \citep[e.g.,][]{PortegiesZwart2010}. In the second limit, we follow the phenomenological results of \citet{Marks2012} which showed initial cluster half-mass radii, $r_{\rm{h}}$ (a reasonable proxy for $r_v$), exhibit a weak dependence on total cluster mass:

\begin{equation}
    \label{eq:Marks}
    r_{\rm{h}} = 0.1^{+0.07}_{-0.04}\,\rm{pc}\Big( \frac{\it{M}_{\rm{cl}}}{\it{M}_{\odot}} \Big)^{0.13 \pm 0.04}.
\end{equation}
As the virial radii given by the Equation \ref{eq:Marks} are a factor of roughly two smaller than the $r_v=1\,$pc assumption, the models adopting this relation are roughly an order of magnitude denser than the $r_v=1\,$pc counterparts. In this case, the various dynamical processes, including stellar-mass BH TDEs occur at an increased rate under the \citet{Marks2012} assumption, as will be discussed further in Section \ref{sec:results}.

For simplicity, we assume zero primordial stellar binaries in all models in this study. Under this assumptions, all TDEs occur as a result of well-understood dynamical processes, with no assumptions required regarding the uncertain properties of primordial stellar binaries in clusters. However, we note that primordial binaries will likely lead to an increase TDE rate \citep[for example, see][which explored the role of primordial binaries in the similar topic of stellar collisions]{Fregeau2007}. In this case, the results of this study may be viewed as a conservative lower limit on the true TDE rate in YSCs. We return to this question in Section \ref{sec:conclusion}.

Finally, to increase the robustness of our results, we run a large number of independent realizations of each set of cluster initial conditions. In total, we produce $3010$ independent models. The complete list of models, including initial conditions and numbers of TDEs, is shown in Table \ref{table:models}.

\vspace{1cm}

\section{Results}
\label{sec:results}

In this section, we show the results of our suite of $N$-body models. In Sections \ref{sec:single_BH_TDEs} and \ref{sec:BBH_TDEs}, we discuss TDEs that occur during single--single and binary--single dynamical encounters, respectively, and compare to simple analytic estimates. In Section \ref{sec:rates}, we examine the properties of TDEs and how these properties vary with both the cluster metallicity and initial density. Finally, in Section \ref{sec:masses}, we estimate the overall rates of TDEs at various redshifts.

\subsection{Single--single TDEs}
\label{sec:single_BH_TDEs}

We first discuss the case of TDEs occuring during single--single encounters between a BH and a MS star. For a given cluster with $N$ total stars and half-mass radius $r_{\rm{h}}$, the rate of TDEs occurring through single--single dynamical encounters can be estimated as

\begin{equation}
    \label{eq:rate_model}
    \Gamma_{\rm{ss}} \approx n_{\star} \Sigma_{\rm{ss}} \sigma_v N_{\rm{BH}}
\end{equation}
where $N_{\rm{BH}}$ is the total number of BHs in the cluster \citep[which, in general, are all found within the half-mass radius due to mass segregation; e.g., ][]{Morscher2015}, $n_{\star} \approx N/r_{\rm{h}}^3$ is the number density, and $\sigma_v$ is the cluster's velocity dispersion.
$\Sigma_{\rm{ss}}$ is the cross section for a single--single TDE, given by

\begin{equation}
    \label{eq:sigma_ss}
    \Sigma_{\rm{ss}} = \pi R_{\rm{TD}}^2 \Bigg[ 1 + \frac{2 G(m_{\rm{BH}}+m_{\star})}{R_{\rm{TD}}\sigma_v^2} \Bigg]
\end{equation}
where $m_{\rm{BH}}$ and $m_{\star}$ are the typical BH and stellar masses and $R_{\rm{TD}}$ is the tidal disruption radius given by Equation \ref{eq:r_TD}. For a cluster with a \citet{Kroupa2001} IMF, we can take $m_{\star} \approx 0.6\,M_{\odot}$. For high-metallicity clusters ($Z\approx Z_{\odot}$), we can take $m_{\rm{BH}} \approx 10\,M_{\odot}$, while for low-metallicity clusters ($Z\lesssim 0.1Z_{\odot}$), $m_{\rm{BH}} \approx 25\,M_{\odot}$ is more appropriate \citep[e.g.,][]{Kremer2020}.

Assuming the cluster is initially in virial equilibirum  ($\sigma_v \approx \sqrt{GM_{\rm{cl}}/r_{\rm{h}}}$,
where $M_{\rm{cl}}\approx m_{\star}N$ is the total cluster mass), assuming all encounters occur in the gravitational focusing regime (the second term in the brackets of Equation \ref{eq:sigma_ss} dominates), and taking $m_{\rm{BH}} \gg m_{\star}$, we can rewrite Equation \ref{eq:rate_model} as

\begin{multline}
    \label{eq:rate_per_cluster}
    \Gamma_{\rm{ss}} \approx 0.2\,\rm{Gyr}^{-1}
    \Big(\frac{\it{M_{\rm{cl}}}}{10^4\,M_{\odot}} \Big)^{1/2}
    \Big(\frac{\it{r}_{\rm{h}}}{1\,\rm{pc}} \Big)^{-5/2} \\ \times \Big(\frac{\it{m}_{\star}}{0.6\,\it{M}_{\odot}} \Big)^{-4/3}
     \Big(\frac{\it{m}_{\rm{BH}}}{10\,M_{\odot}} \Big)^{4/3}
     \Big(\frac{\it{R}_{\star}}{0.6\,\it{R}_{\odot}} \Big)
     \Big(\frac{\it{N}_{\rm{BH}}}{10} \Big).
\end{multline}
For a \citet{Kroupa2001} IMF, we expect roughly $10^{-3}$ BHs to form per star, giving us $N_{\rm{BH}}\approx 10^{-3} N$. In this case, we obtain the analytic scaling

\begin{equation}
    \label{eq:low}
    \Gamma_{\rm{ss}} \approx 0.2\,\rm{Gyr}^{-1} \Big(\frac{\it{M}_{\rm{cl}}}{10^4\,\it{M}_{\odot}} \Big)^{3/2}.
\end{equation}

\begin{figure}
\begin{center}
\includegraphics[width=\linewidth]{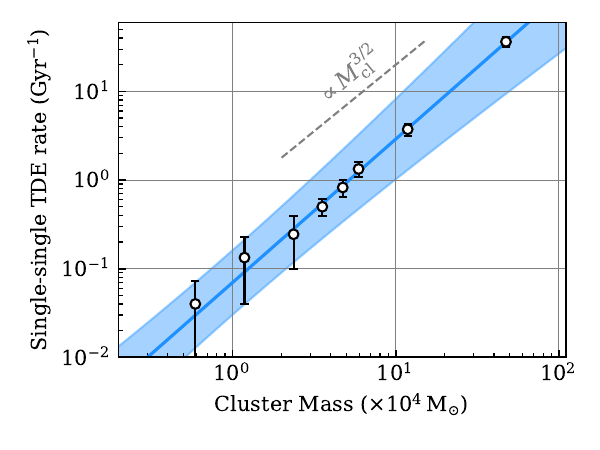}
\caption{\footnotesize \label{fig:rates} TDE rate per cluster as a function of initial cluster mass using models \texttt{a-h} in Table \ref{table:models} (assuming $r_h=1\,$pc). Open circles denote rates computed from the suite of $N$-body models and the solid blue curve shows the best-fit relation of Equation \ref{eq:bf_singlesingle}. The shaded blue region denotes the $90\%$ confidence interval from the least squares fit. The dashed gray line shows the $\propto M_{\rm{cl}}^{3/2}$ analytic scaling from Equation \ref{eq:low}.}
\end{center}
\end{figure}

In Figure \ref{fig:rates} we show as open circles the rates of such encounters as a function of cluster mass as determined from our $N$-body models. To isolate specifically the effect of cluster mass, we utilize only the first 8 sets of simulations listed in Table \ref{table:models} (models \texttt{a-h}), which have fixed metallicity ($Z_{\odot}$) and virial radius ($1\,$pc). To calculate the model rates, we simply count the total number of single--single TDEs in all models of a given cluster mass, then divide by the total number of models and by the total integration time for the given cluster mass (see Table \ref{table:models}). Error bars denote $2\sigma$ from the mean, assuming a Poisson distribution.

From a least squares fit, we find these data are best fit by the power-law relation

\begin{equation}
    \label{eq:bf_singlesingle}
    \Gamma_{\rm{ss}}^{\rm{model}} = 0.07 \pm 0.01\,\rm{Gyr}^{-1} \Big(\frac{\it{M}_{\rm{cl}}}{10^4\,M_{\odot}} \Big)^{1.6 \pm 0.16}.
\end{equation}
This fit is shown as the blue curve in Figure \ref{fig:rates}, with the blue bands denoting the $90\%$ confidence interval from the least squares fit.
For comparison, we show as the dashed gray line in Figure \ref{fig:rates} the $\Gamma \propto M_{\rm{cl}}^{3/2}$ scaling relation derived from the simple analytic estimate in Equation \ref{eq:low}.

In Equation \ref{eq:low}, we have assumed a constant $r_{\rm{h}}\approx1\,$pc is typical for all YSCs \citep[see, e.g.,][]{PortegiesZwart2010}. Alternatively, as discussed in Section \ref{sec:methods}, $r_{\rm{h}}<1\,$pc may be more appropriate and furthermore, $r_{\rm{h}}$ may exhibit a weak dependence on total cluster mass. To explore the possibility, we ran an additional set of models (group \texttt{l} in Table \ref{table:models}) with initial $r_v=0.42\,$pc, reflective of the $r_{\rm{h}}-M_{\rm{cl}}$ relation of \citet{Marks2012}. Given the phenomenological relation from \citet{Marks2012} predicts cluster radii a factor of roughly 2 lower than the $r_{\rm{h}}=1\,$pc assumption, the following rate, $\Gamma_{\rm{ss}}^{\rm{phenom}}$, is higher than that estimated from Equation \ref{eq:bf_singlesingle}.

Combining the phenomenological $r_{\rm{h}}-M_{\rm{cl}}$ relation of \citet{Marks2012} (Equation \ref{eq:Marks}) with Equations \ref{eq:rate_model} and \ref{eq:sigma_ss}, we expect $\Gamma_{\rm{ss}}^{\rm{phenom}} \propto M_{\rm{cl}}^{1.175}$. Scaling to the rate identified from the models in group \texttt{l}, we can then write:

\begin{equation}
    \label{eq:opt_singlesingle}
    \Gamma_{\rm{ss}}^{\rm{phenom}} \approx 0.63\,\rm{Gyr}^{-1} \Big(\frac{\it{M}_{\rm{cl}}}{10^4\,\it{M}_{\odot}} \Big)^{1.175}.
\end{equation}

\subsection{Binary-mediated TDEs}
\label{sec:BBH_TDEs}

In addition to TDEs occurring through single--single encounters, TDEs may also take place during binary-mediated resonant encounters involving at least one BH and one star. As discussed in Section \ref{sec:methods}, we do not include stellar binaries in this study. The only binaries formed are BH binaries assembled through three-body encounters \citep[for simplicity, three-body binary formation is allowed \textit{only} for BHs in our models; e.g.,][]{Morscher2015}. In this case, the binary-mediated TDEs discussed in this subsection are those that occur specifically during binary--single resonant encounters between a BBH and a single MS star.

Using a similar calculation to that performed for the single--single rate estimate, the rate of TDEs during BBH--star binary--single encounters can be written as

\begin{equation}
    \label{eq:TDE,BBH}
    \Gamma_{\rm{bs}} \approx n_{\star}\,\Sigma_{\rm{bs}} \, \sigma_v \, N_{\rm{BBH}} \, P_{\rm{TD}}.
\end{equation}
Here $\Sigma_{\rm{bs}}$ is the cross section for binary--single encounters

\begin{equation}
    \label{eq:sigma}
    \Sigma_{\rm{bs}} = \pi a_{\rm{BBH}}^2 \Bigg[ 1 + \frac{2 G(m_{\rm{BBH}}+m_{\star})}{a_{\rm{BBH}}\sigma_v^2} \Bigg]
\end{equation}
where $a_{\rm{BBH}}$ is the BBH semi-major axis and $m_{\rm{BBH}}\approx 2 m_{\rm{BH}}$ is the mass of the BBH. $N_{\rm{BBH}}$ is the total number of BBHs in the cluster. As shown in a number of recent analyses \citep[e.g.,][]{Morscher2015,Chatterjee2017a,Banerjee2018b}, $N_{\rm{BBH}}$ is expected to be roughly independent of the total number of BHs in the cluster as well as the cluster's total mass, such that the total number of dynamically-formed BBHs present at any given time never exceeds a few. Here we assume $N_{\rm{BBH}}\approx2$.

Finally, $P_{\rm{TD}}$ is the probability that a given BBH--star resonant encounter leads to a TDE. We follow the results of \citet{Darbha2018}, which showed that for asymmetric mass ratio binary-single encounters, $P_{\rm{TD}}$ is roughly proportional to $R_{\rm{TD}}/a_{\rm{BBH}}$. Note that this same scaling is found in the equal mass case \citep[e.g.,][]{Samsing2017,Samsing2018}. Here we take $P_{\rm{TD}} = 2 R_{\rm{TD}}/a_{\rm{BBH}}$ as in \citet{Samsing2019}. We can then rewrite Eq. \ref{eq:TDE,BBH} as

\begin{equation}
    \label{eq:BBHrate_per_cluster}
    \Gamma_{\rm{bs}} \approx 0.3\,\rm{Gyr}^{-1}
    \Big(\frac{\it{M_{\rm{cl}}}}{10^4\,M_{\odot}} \Big)^{1/2}
\end{equation}
where, as before, we have assumed $r_{\rm{h}}\approx1\,$pc, $m_{\rm{BH}}\approx10\,M_{\odot}$, $m_{\star}\approx0.6\,M_{\odot}$, and $R_{\star}\approx0.6\,R_{\odot}$, independent of the cluster mass.

Comparing to Equation \ref{eq:low}, we find $\Gamma_{\rm{bs}}/\Gamma_{\rm{ss}} \propto N_{\rm{BH}}^{-1} \propto M_{\rm{cl}}^{-1}$. Thus, lower-mass clusters with fewer BHs feature a higher BBH TDE rate per BH (and therefore also per star) compared to higher-mass clusters with more BHs.
Thus, given that lower-mass clusters also dominate by number the overall cluster mass function, they are the ideal environment to find BBH TDEs. We return to the question of overall and relative rates in Section \ref{sec:rates}. 

\begin{figure}
\begin{center}
\includegraphics[width=\linewidth]{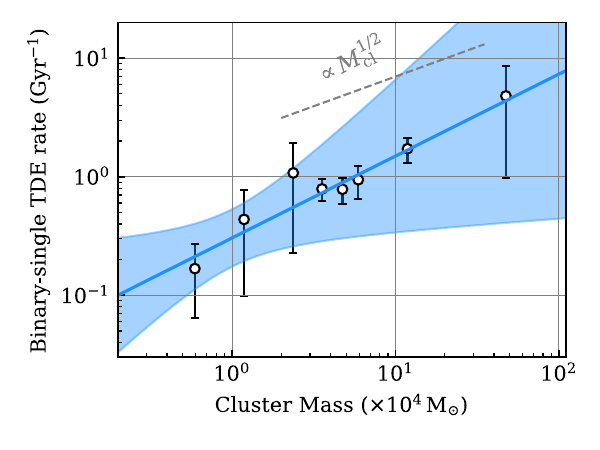}
\caption{\footnotesize \label{fig:ratesBBH} Same as Figure \ref{fig:rates} but for TDEs occurring during binary--single encounters. Here the blue curve shows the best-fit relation of Equation \ref{eq:opt_binarysingle} (as in Figure \ref{fig:rates}, the shaded blue region denotes the $90\%$ confidence interval from the least squares fit) and the dashed gray line shows the $\propto M_{\rm{cl}}^{1/2}$ analytic scaling from Equation \ref{eq:BBHrate_per_cluster}.}
\end{center}
\end{figure}

In Figure \ref{fig:ratesBBH} we show the rates of these events as a function of cluster mass as determined from our $N$-body models. As in Figure \ref{fig:rates}, the rate is computed as the total number of binary--single TDEs in all models of a given cluster mass, divided by the total number of models and by the time duration, $\Delta t$. Unlike in the single--single case, where the necessary dynamical encounters begin immediately, in the binary--single case we must first wait for BHs to mass-segregate to the center so that the target BBHs can form through three-body encounters. For each model, we define this timescale, $t_{\rm{3bb}}$, simply as the moment the first BBH forms in that model. Then, $\Delta t = t_{\rm{max}}-t_{\rm{3bb}}$. Typically, $t_{\rm{3bb}}$ is of order $100\,$Myr \citep[see also][]{Sigurdsson1993,Morscher2015}.

Again performing a least squares fit, we find these data are best fit by the power-law relation

\begin{equation}
    \label{eq:bf_binarysingle}
    \Gamma_{\rm{bs}}^{\rm{model}} = 0.69\pm 0.36\,\rm{Gyr}^{-1} \Big(\frac{\it{M}_{\rm{cl}}}{10^4\,M_{\odot}} \Big)^{0.3 \pm 0.09},
\end{equation}
which can be compared with the simple analytic estimate of Eq. \ref{eq:BBHrate_per_cluster}. As with the single--single case, we find the analytic estimate recovers reasonably well the rate inferred from the $N$-body modelling.

As before, we can also estimate a phenomenological rate, $\Gamma_{\rm{bs}}^{\rm{phenom}}$, assuming the relation of \citet{Marks2012}. Combining Equations \ref{eq:Marks} and \ref{eq:BBHrate_per_cluster}, we expect $\Gamma_{\rm{bs}}^{\rm{phenom}} \propto M_{\rm{cl}}^{0.175}$. Again normalizing to the binary--single TDE rate estimated in the models form group \texttt{l} in Table \ref{table:models}, we can write

\begin{equation}
    \label{eq:opt_binarysingle}
    \Gamma_{\rm{bs}}^{\rm{phenom}} \approx 4.4\,\rm{Gyr}^{-1} \Big(\frac{\it{M}_{\rm{cl}}}{10^4\,M_{\odot}} \Big)^{0.125}.
\end{equation}

Because we assume that BH binaries form exclusively through three-body encounters, Equations \ref{eq:bf_binarysingle} and \ref{eq:opt_binarysingle} are relevant only for those clusters with at least three BHs at birth, so that at least one BBH can form. Again, assuming roughly 1 BH forms per 1000 stars \citep{Kroupa2001}, this requires $N \gtrsim 3000$ or $M_{\rm{cl}}\gtrsim 2000\,M_{\odot}$. For clusters with $M_{\rm{cl}}\lesssim 2000\,M_{\odot}$, Equations \ref{eq:bf_binarysingle} and \ref{eq:opt_binarysingle} are no longer applicable and the binary--single TDE rate is zero.
We return to this point in Section \ref{sec:rates} when we compute the total TDE rate by integrating over the full cluster mass function.

\subsubsection{Binary BH orbital separations}

\begin{figure*}
\begin{center}
\includegraphics[width=0.7\linewidth]{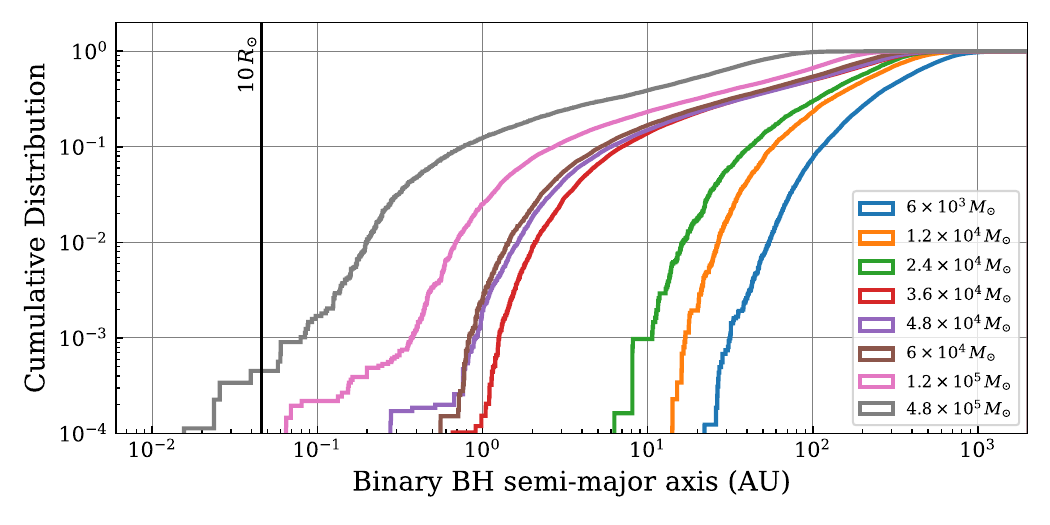}
\caption{\footnotesize \label{fig:BBH_Porb} Cumulative distribution of semi-major axis for all BBHs identified in models \texttt{a-h}. The various colors denote different cluster masses. As shown, more massive clusters host, on average, more compact BBHs. The solid black line shows the characteristic orbital separation at which the BH binary companion may interrupt the TDE lightcurve, as described in the text.}
\end{center}
\end{figure*}

Given that the binary--single TDE rate is expected to be roughly independent of the binary orbital separation, the distribution of $a_{\rm{BBH}}$ for BBHs that undergo TDEs is expected to follow the semi-major axis for all BBHs in a cluster of a given mass \citep[e.g.,][]{Samsing2019,KremerD'OrazioSamsing2019}. In Figure \ref{fig:BBH_Porb}, we show the distribution of $a_{\rm{BBH}}$ for all BBHs found in models \texttt{a-h} (various initial masses, but fixed $r_v$ and $Z$). These distributions are determined by two primary physical processes:

\textit{Three-body formation:} The maximum semi-major axis of a binary formed through a three-body encounter is determined by the hard-soft boundary $a_{\rm{HS}} \propto m_{\rm{BH}}/\sigma_v^2$ \citep[e.g.,][]{Morscher2015}. Assuming $\sigma_v^2 \propto M_{\rm{cl}}$, we expect, in general, more massive clusters will produce more compact BBHs compared to lower mass clusters (assuming a fixed cluster virial radius).

\textit{Ejection from dynamical recoil:} Once a BBH is formed, it will (on average) harden through subsequent binary-mediated encounters with other BHs and stars in the cluster core \citep[e.g.,][]{Sigurdsson1993,Morscher2015,Rodriguez2016a}. Following a dynamical encounter, the BBH will receive a dynamical recoil kick with a magnitude comparable to the BBH orbital velocity, $v_{\rm{recoil}}^2 \propto a_{\rm{BBH}}^{-1}$. Thus, as a BBH hardens, it attains increasingly large dynamical recoil kicks. Eventually, $v_{\rm{recoil}}$ is sufficiently large for the binary to be ejected from the cluster. This is set by the cluster's escape velocity $v_{\rm{esc}}^2 \propto M_{\rm{cl}}/r_{\rm{h}}$. As a result, in lower-mass clusters, BBHs will be dynamically ejected before they can harden as far as is possible in higher-mass clusters.

As a consequence of these two processes, we expect the BBH semi-major axis distribution to shift toward lower values in increasingly massive clusters. Indeed, this is shown in Figure \ref{fig:BBH_Porb}.

One exciting possibility proposed in \citet{Lopez2018,Samsing2019,KremerD'OrazioSamsing2019} is that these binary-mediated TDEs may be used to indirectly probe properties of the underlying BBH population, if the corresponding electromagnetic (EM) signal can be detected. The basic idea is the second BH produces breaks in the lightcurve on timescale comparable to the BBH orbital period. This idea has been illustrated in the supermassive BH (SMBH) regime using numerical techniques \citep[e.g.,][]{Liu2009,Coughlin2017}, and one SMBH candidate has proposed (TDE J1201+30), from which the authors were able to constrain the SMBH binary orbital period \citep{Liu2014}. This process likely requires $a_{\rm{BBH}}$ to be comparable to the disk radius, which in turn will be comparable to the tidal disruption radius. For reference, we show $r_{\rm{TD}}\approx 10\,R_{\odot}$ as a solid black line in Figure \ref{fig:BBH_Porb} (see Equation \ref{eq:r_TD}). As shown in the figure, this process is likely only possible in the most massive clusters explored here ($M_{\rm{cl}} \gtrsim 10^5\,M_{\odot}$). Even for our massive cluster simulations, we find that only $\approx0.1\%$ of all BBHs meet this criterion. Thus, we conclude that the presence of a second BH is unlikely to significantly affect the TDE dynamics and subsequent lightcurve evolution.

Although it appears this possibility is not relevant in typical YSCs, we can speculate that more massive clusters such as nuclear star clusters with masses of $10^7\,M_{\odot}$ or larger may be ideal environments for this processes, given that more massive clusters should be able to host even more compact BBHs. We reserve a more detailed study of this possibility for future study, and direct the reader to \citet{Fragione2020} for a discussion of TDEs in the nuclear star cluster regime.

\subsection{Estimating the total event rate}
\label{sec:rates}

The functional form for the initial mass function of YSCs is expected to be well-represented by a power-law distribution \citep[e.g.,][]{LadaLada2003} with a possible exponential truncation above cluster masses of roughly $M_{\rm{cut}}\approx 10^6\,M_{\odot}$ \citep[e.g.,][]{PortegiesZwart2010}:

\begin{equation}
    \label{eq:Lada}
    \frac{dN}{dM_{\rm{cl}}} \propto M_{\rm{cl}}^{-2} \exp (-M_{\rm{cl}}/M_{\rm{cut}}).
\end{equation}
As we are interested here primarily in the low-mass tail of the mass function ($\lesssim 10^5\,M_{\odot}$), the specific value of $M_{\rm{cut}}$ is not relevant to this study.

We can then compute the total TDE rate from a realistic population of YSCs in Milky Way-like galaxies by integrating the rate per cluster (Eq. \ref{eq:rate_per_cluster}) over the cluster mass function:

\begin{equation}
    \label{eq:integral}
    \Gamma_{\rm{tot}} = \int_{M_{\rm{low}}}^{M_{\rm{high}}} \frac{\Gamma_{\rm{cl}}}{M_{\rm{cl}}} \, \frac{dN}{dM_{\rm{cl}}} \, \Delta t \,  \rho_{\rm{SF}} \, f_{\rm{SF}}\,dM_{\rm{cl}}.
\end{equation}
The integration limits represent the assumed range in cluster masses; we assume  $M_{\rm{low}}=100\,M_{\odot}$ and $M_{\rm{high}}=10^5\,M_{\odot}$ \citep{LadaLada2003}. By definition, in order for TDEs to occur through the binary--single channel discussed in Section \ref{sec:BBH_TDEs}, BBHs must have formed in the cluster. As discussed in Section \ref{sec:BBH_TDEs}, because we assume that BBHs form exclusively through three-body encounters, this requires at least three BHs be present in the cluster at birth, which in turn requires $M_{\rm{cl}}\gtrsim 2000\,M_{\odot}$. Thus, to compute the rate of TDEs occurring through binary--single encounters, we use $M_{\rm{low}}=2000\,M_{\odot}$ (keeping the overall normalization of Equation \ref{eq:Lada} the same as before). Note that this mass requirement automatically takes care of the additional requirement that the cluster not dissolve before the first BBHs begin to form. 

$\Delta t$ is the cluster disruption timescale, $t_{\rm{dis}}$, which depends upon the location of the cluster in its host galaxy's tidal field, as well as on more complex phenomena such as, e.g., tidal shocks and interactions with giant molecular clouds as mentioned in Section \ref{sec:methods}. Here we adopt the following relation from \citet{Lamers2005}:

\begin{equation}
    t_{\rm{dis}} \simeq 810 \Bigg( \frac{M_{\rm{cl}}}{10^4\,M_{\odot}} \Bigg)^{0.62} \Bigg( \frac{\rho_{\rm{amb}}}{M_{\odot}\, \rm{pc}^{-3}} \Bigg)^{-0.5} \rm{Myr}
\end{equation}
where $\rho_{\rm{amb}} \approx M_{\odot}\, \rm{pc}^{-3}$ is the assumed local ambient density.
Again, as we are specifically interested in young clusters with ages less than roughly $500\,$Myr, we take

\begin{equation}
    \Delta t = \min \Big( t_{\rm{dis}}, 500\,\rm{Myr} \Big)
\end{equation}

For TDEs occurring during binary--single encounters, we must also incorporate the timescale for BBH formation to begin. In Section \ref{sec:BBH_TDEs}, we computed this timescale directly from the models. Here, for simplicity (and as motivated by the results from the models), we assume BBH formation occurs after roughly $100\,$Myr for all cluster masses \citep[see also, e.g.,][]{Sigurdsson1993}. In this case, for binary--single TDEs, $\Delta t = \min(t_{\rm{dis}}, 500\,\rm{Myr}) - 100\,\rm{Myr}$.

$\rho_{\rm{SF}}$ is the assumed cosmological density of star formation rate (SFR). We adopt the SFR of \citet{HopkinsBeacom2006}. Specifically, this study finds $\rho_{\rm{SF}}/[M_{\odot} \, \rm{yr}^{-1}\,\rm{Mpc}^{-3}] =1.5\times10^{-2}$ , $0.1$, and $0.2$ at redshift $z=0, 1$, and $2.5$ (peak star formation), respectively. $f_{\rm{SF}}$ is the fraction of the star formation rate assumed to occur in star clusters. For low-mass clusters ($M_{\rm{cl}} < 10^5\,{\odot}$) we assume $f_{\rm{SF}}=0.8$ \citep{LadaLada2003}.

Finally, $\Gamma$ is the TDE rate per cluster of a given mass, $M_{\rm{cl}}$. For this we adopt the scaling relations derived in Section \ref{sec:results} (Equations \ref{eq:bf_singlesingle} and \ref{eq:bf_binarysingle}, for the single--single and binary--single cases, respectively, assuming constant $r_{\rm{h}}=1\,$pc). Additionally, we use Equations \ref{eq:opt_singlesingle} and \ref{eq:opt_binarysingle} to compute the rates assuming higher density YSCs as in the phenomenological fits of \citet{Marks2012}.

We present in Table \ref{table:rates} the rate estimates obtained by integrating Equation \ref{eq:integral}. We find that the binary--single channel dominates over the single--single channel by a factor of roughly a few to ten, depending upon the assumptions made regarding $r_{\rm{h}}$. If we adopt the more conservative choice of constant $r_{\rm{h}}=1\,$pc, we estimate a combined TDE rate of roughly $20\,\rm{Gpc}^{-3}\,\rm{yr}^{-1}$ in the local universe. Adopting the phenomenological assumption from \citet{Marks2012}, we find a combined TDE rate of roughly $200\,\rm{Gpc}^{-3}\,\rm{yr}^{-1}$. 

For reference, \citet{Kremer2019c} predicted a TDE rate of roughly $10\,\rm{Gpc}^{-3}\,\rm{yr}^{-1}$ for old globular clusters. In the more massive nuclear star cluster regime, \citet{Fragione2020} predicted a stellar-mass BH TDE rate of roughly $10^{-7}-10^{-6}\,\rm{yr}^{-1}$ per galaxy. Assuming a galactic density of roughly $10^{-2}\,\rm{Mpc}^{-3}$, this corresponds to a rate of roughly $1-10\,\rm{Gpc}^{-3}\,\rm{yr}^{-1}$ in the local universe. Thus, we conclude YSCs may dominate the overall stellar-mass BH TDE rate in the local universe by a factor of a few to more than an order of magnitude compared to more massive clusters.

\begin{deluxetable}{c|c|c|c}
\tabletypesize{\normalsize}
\tablewidth{0pt}
\tablecaption{Volumetric TDE rates \label{table:rates}}
\tablehead{
    \colhead{\hspace{.5cm}$r_{\rm{h}}$ prescription} &
	\colhead{\hspace{.5cm}$z=0$} &
	\colhead{\hspace{.5cm}$z=1$} &
	\colhead{\hspace{.5cm}$z=2.5$}\\
	\colhead{} &
	\multicolumn{3}{c}{($\rm{Gpc}^{-3}\,\rm{yr}^{-1}$)}
}
\startdata
\multicolumn{4}{c}{Single--single TDE rate:}\\
\hline
$r_{\rm{h}}=1\,$pc & 1.1 & 7.1 & 12.1\\
\small{Marks \& Kroupa 2012} & 27.6 & 184.3 & 313.3\\
\hline
\multicolumn{4}{c}{Binary--single TDE rate:}\\
\hline
$r_{\rm{h}}=1\,$pc & 19.3 & 128.4 & 218.3\\
\small{Marks \& Kroupa 2012} & 157.7 & 1051.2 & 1787.0\\
\enddata
\tablecomments{ \small Volumetric event rates of TDEs occurring through both single-single and binary-single encounters in YSCs in the local universe ($z=0$), at $z=1$, and at peak star formation ($z\approx2.5$). We show both rates calculated assuming constant cluster half-light radii, $r_{\rm{h}}=1\,$pc and assuming the $r_{\rm{h}}-M_{\rm{cl}}$ relation of \citet{Marks2012} (Equation \ref{eq:Marks}).}
\end{deluxetable}

\subsection{BH mass function of TDEs}
\label{sec:masses}

\begin{figure}
\begin{center}
\includegraphics[width=\linewidth]{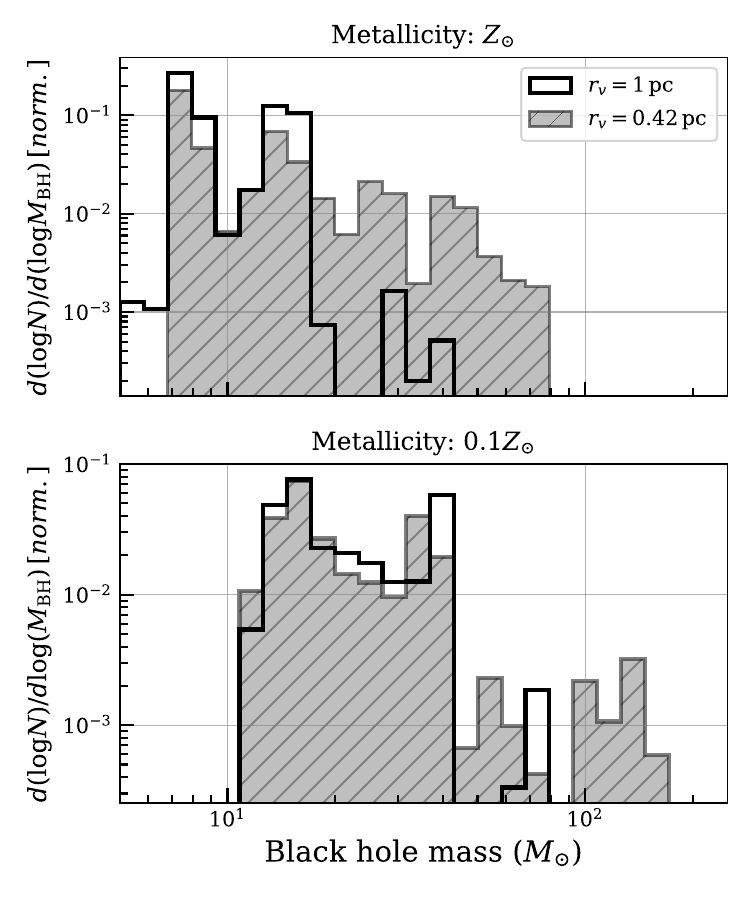}
\caption{\footnotesize \label{fig:masses} Normalized distribution of BH masses for all TDEs occurring in the various cluster models. In the top (bottom) panel, we show mass distributions for $Z_{\odot}$ ($0.1Z_{\odot}$) models. The solid black and hatched gray histograms denote models adopting a constant $r_v=1\,$pc and the $r_{\rm{h}}-M_{\rm{cl}}$ relation from \citet{Marks2012}, respectively.}
\end{center}
\end{figure}

In the previous subsections, we have explored specifically models \texttt{a}-\texttt{h} of Table \ref{table:models} which assume solar metallicity, reflective of YSCs born recently in the local universe. However, for YSCs found at higher redshifts, assuming a lower metallicity is more appropriate. Given the overall TDE rate be substantially higher at high redshift (see Section \ref{sec:rates}), a careful investigation of metallicity is warranted.

To explore the effect of cluster metallicity on TDE properties, we have run several additional sets of models with $Z=0.1Z_{\odot}$. In Figure \ref{fig:masses}, we show the distribution of BH masses that undergo TDEs. The black and hatched gray histograms show BH masses found in models assuming $r_{\rm{h}}=1\,$pc and the \citet{Marks2012} $r_{\rm{h}}-M_{\rm{cl}}$ relation, respectively. In the top (bottom) panel we show the results for models assuming solar ($10\%$ solar) metallicity.

The first general feature we see is that higher metallicity clusters yield lower mass BH TDEs. This is anticipated: at higher metallicity, stellar winds are expected to lead to increased mass loss prior to stellar core collapse, which in turn is expected to reduce the mass of the BH ultimately formed \citep[e.g.,][]{Vink2001,Fryer2012,Belczynski2016b}. For the $r_v=1\,$pc models, we find median BH TDE masses of $13\,M_{\odot}$ and $27\,M_{\odot}$ for $Z_{\odot}$ and $0.1Z_{\odot}$, respectively.

We also see that, for a given metallicity, increasing the initial cluster density yields an extended tail in the upper part of the BH mass spectrum. This high-mass tail is populated by BHs formed through stellar collisions, which occur at an increased rate in higher density clusters. A number of recent analyses \citep{Spera2017,DiCarlo2019,DiCarlo2020,Kremer2020c} have demonstrated that dynamically-mediated stellar collisions occurring within the first roughly $5\,$Myr of cluster evolution (before formation of BHs) may lead to formation of massive stars that may ultimately collapse to form high-mass BHs. In particular, this process may permit formation of BHs with masses occupying the pair-instability mass gap from roughly $40-120\,M_{\odot}$ expected to arise through (pulsational) pair-instability supernovae \citep[e.g.,][]{Belczynski2016b,Woosley2017,Spera2017}. Additionally, this process may be closely related to collisional runways which have been touted as a potential formation mechanism for intermediate-mass BHs (IMBHs) with masses in excess of roughly $100\,M_{\odot}$ \citep[e.g.,][]{PortegiesZwart2004,Gurkan2006,Giersz2015,Mapelli2016}. Here, we adopt the same prescriptions implemented in \citet{Kremer2020c} to treat stellar collisions and the subsequent evolution of the collision products, as described in Section \ref{sec:methods}. We also assume the maximum BH mass formed through single-star evolution (i.e., unaffected by any dynamical processes) is $40.5\,M_{\odot}$, as determined by our (pulsational) pair-instability supernova treatment \citep[see][for detail]{Belczynski2016b}.

In our $Z_{\odot}$ and $0.1Z_{\odot}$
models that adopt the \citet{Marks2012} $r_{\rm{h}}$ relation, we find roughly $10\%$ of all TDEs have BH masses within the assumed pair-instability gap. In the $0.1Z_{\odot}$ models specifically, we find an additional $5\%$ of TDEs occur with a BH mass in excess of $120\,M_{\odot}$ (the assumed upper limit to the pair-instability gap). These fractions are consistent with the rates of formation of massive BHs shown in previous studies of YSCs \citep{DiCarlo2019,DiCarlo2020}. Given the cosmological rates predicted in Section \ref{sec:rates}, these mass-gap and IMBH TDEs may constitute a non-negligible fraction of all TDEs occurring in YSCs. Furthermore, these TDEs may provide a potentially novel way to probe the formation of pair-instability gap BHs, similar to the recent LIGO/Virgo detection GW190521 \citep{LIGO2020a,LIGO2020b}.

Finally, comparing models \texttt{i,j,k} with models \texttt{e,f,g}, we see that lower metallicity clusters exhibit a moderate increase (a factor of roughly 1.4) in the total number of TDEs occurring through both single--single and binary--single encounters. This slight increase in the rate is anticipated: as shown in Equation \ref{eq:rate_per_cluster}, the TDE rate scales with the BH mass, $m_{\rm{BH}}$, through the influence of gravitational focusing and through tidal disruption radius calculation (Equation \ref{eq:r_TD}). 
Thus, we expect that metallicity leads to a moderate difference in the TDE rate for a given cluster mass.

\section{Electromagnetic signatures}
\label{sec:EMsignatures}

In the previous section, we have shown that stellar-mass BH TDEs should be plentiful in YSCs. We now examine possible electromagnetic signatures of these events, building upon previous work on this subject \citep[e.g.,][]{Perets2016,Kremer2019c,Lopez2018,Fragione2019}.

\subsection{Characteristic timescales and luminosities}
\label{sec:EMfeatures}

\begin{figure*}
\begin{center}
\includegraphics[width=0.9\linewidth]{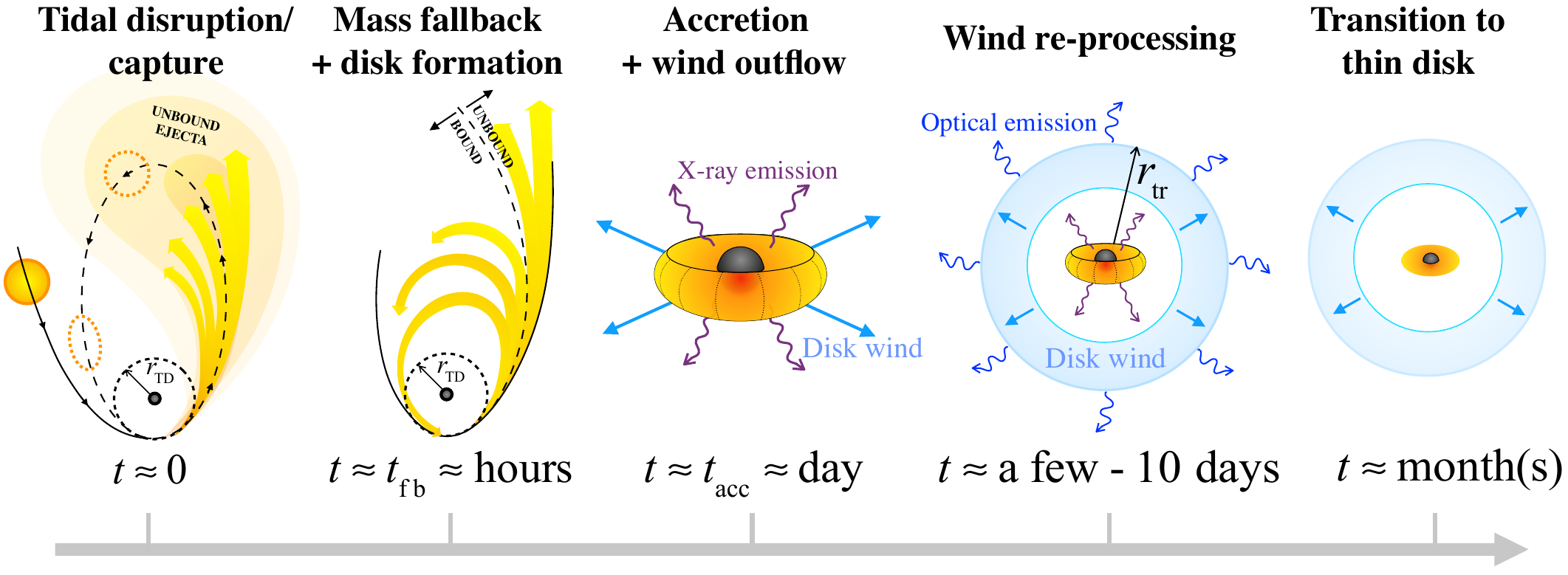}
\caption{\footnotesize \label{fig:TDE} Schematic illustration of a stellar-mass BH tidal disruption event including disk formation and evolution in time. From left to right, we show: (1) Tidal disruption of the star, allowing for a possible initial partial disruption that unbinds a small fraction of stellar mass while the star is tidally captured into an elliptical orbit, (2) Fallback of bound material to pericenter, (3) Rise time for X-ray emission ($L_X \gtrsim 10^{44}\,\rm{erg\,s}^{-1}$; Equation \ref{eq:Xray}) through viscous accretion onto the BH, (4) Re-processing of the X-ray emission by disk wind at the trapping radius leads to bright optical emission ($L_{\rm{opt}} \approx 10^{41}-10^{44}\,\rm{erg\,s}^{-1}$), (5) Transition to thin disk and prompt drop in $\dot{M}$ and luminosity.
}
\end{center}
\end{figure*}

Following the disruption of a star of mass $m_{\star}$ and radius $R_{\star}$, the timescale, $t_{\rm{fb}}$ for (bound) orbiting material to fallback to the disruption point, $R_{\rm{TD}}$ is given by:

\begin{multline}
    t_{\rm{fb}} = \frac{\pi R_{\rm{TD}}^3}{\sqrt{2Gm_{\rm{BH}}R_{\star}^3}} \\
    \approx 1.1\times10^4\,\rm{s} \Big( \frac{\it{m}_{\star}}{\it{M}_{\odot}} \Big)^{-1} \Big( \frac{\it{R}_{\star}}{\it{R}_{\odot}} \Big)^{3/2} \Big( \frac{\it{m}_{\rm{BH}}}{10\,\it{M}_{\odot}} \Big)^{1/2}
\end{multline}
\citep[e.g.,][]{Perets2016}.

For simplicity, we assume a disk is formed promptly at radius $r_d \simeq R_{\rm TD}$ and take $t_{\rm{fb}}$ as the characteristic timescale for disk formation. This is motivated by the fact that the orbits of the bound debris are only weakly eccentric. Once a disk is formed, the timescale for the debris to accrete is set by the viscous timescale \citep{ShakuraSunyaev1973}. For a thick disk, with disk height ratio $h = H/r_d$ (where $H$ is the disk scale height), the viscous accretion timescale is

\begin{multline}
    t_{\rm{acc}} \approx \Big[ h^2\alpha \Omega_{\rm{K}}(r_d) \Big]^{-1} \\
    \approx 6\times10^4\,\rm{s} \,\Big( \frac{\textit{h}}{0.5} \Big)^{-2} \Big( \frac{\alpha}{0.1} \Big)^{-1} \Big( \frac{\it{m}_{\star}}{\it{M}_{\odot}} \Big)^{-1/2}  \Big( \frac{\it{R}_{\star}}{\it{R}_{\odot}} \Big)^{3/2}
\end{multline}
Because $t_{\rm{acc}} > t_{\rm{fb}}$, the subsequent lightcurve evolution is viscosity-driven (i.e., determined by accretion timescale). This marks one key difference from supermassive BH TDEs, where $t_{\rm{acc}} \ll t_{\rm{fb}}$ and thus, the disk evolution is likely dominated by the fallback and the accretion rate is believed to follow the standard $t^{-5/3}$ power-law \citep[e.g.,][]{Rees1988,Phinney1989,EvansKochanek1989}.

For viscosity-driven accretion, assuming roughly half of the stellar material is bound to the BH following the TDE, the peak accretion rate can be approximated as
\begin{multline}
    \dot{M}_p \approx \frac{m_{\star}/2}{t_{\rm{acc}}} \\ \approx 2.6 \times 10^2\,M_{\odot} \rm{yr}^{-1} \,\Big( \frac{\textit{h}}{0.5} \Big)^{2} \Big( \frac{\alpha}{0.1} \Big) \Big( \frac{\textit{m}_{\star}}{\it{M}_{\odot}} \Big)^{3/2}  \Big( \frac{\it{R}_{\star}}{\it{R}_{\odot}} \Big)^{-3/2}
\end{multline}
The maximum possible luminosity is $L_{\rm{max}} \sim \epsilon \dot{M}_p c^2$, where $\epsilon\sim 0.1$ is the accretion efficiency near the innermost stable circular orbit (ISCO). This case corresponds to the most efficient energy release (possibly when a jet is formed), and we estimate $L_{\rm{max}} \sim 10^{48}\,\rm{erg\,s}^{-1}$ for typical TDE parameters. As pointed out in \citet{Perets2016}, in this extreme case the TDE may power an ultra-long gamma ray burst \citep[e.g.,][]{Gendre2013,Levan2014}.

In the more widely accepted adiabatic inflow–outflow (ADIOS) model \citep{BlandfordBegelman1999}, the mass inflow of a super-Eddington disk onto a BH is non-conservative and only a small fraction of the mass supplied at large radii is actually accreted. The accretion rate is expected to be reduced by factor $(10r_g/r_d)^{s}$, where $r_g = G m_{\rm{BH}}/c^2$ is the BH gravitational radius, $r_d$ is the disk radius, and the power-law index $s\in (0, 1)$.
For the most pessimistic case of $s=1$, we can estimate the accretion luminosity of the inner disk, a significant fraction of which should be observable in the X-ray band for favorable viewing angles close to face-on,

\begin{multline}
    \label{eq:Xray}
    L_{\rm min} \approx \epsilon \Big( \frac{10r_g}{r_d} \Big) \dot{M}_p c^2\\
    \approx 1.5 \times 10^{44}\,\rm{erg\,s}^{-1} \,\Big( \frac{\textit{m}_{\star}}{\it{M}_{\odot}} \Big)^{7/6} \Big( \frac{\textit{m}_{\rm{BH}}}{10\,\it{M}_{\odot}} \Big)^{4/3} \Big( \frac{\it{R}_{\star}}{\it{R}_{\odot}} \Big)^{-5/2}
\end{multline}
where we have once again assumed $h=0.5$ and $\alpha=0.1$. For $0<s<1$, we expect the accretion power to be somewhere in between $10^{44}$ and $10^{48}\rm\, erg\, s^{-1}$. For instance, based on numerical simulations of adiabatic accretion flows, \citet{Yuan12_ADIOS} argued for $s\approx 0.5$ which corresponds to $L\sim 10^{46}\rm\, erg\, s^{-1}$.

At later time $t\gg t_{\rm acc}$, the disk radius increases as $r_{d}\propto t^{2/3}$ due to viscous spreading, the mass inflow rate drops as $\dot{M}\propto t^{-2(s+2)/3}$, and the BH accretion rate drops as $\dot{M}_{\rm BH}\propto t^{-4(s+1)/3}$ \citep[e.g.,][]{kumar08_ADAF_evolution}. Thus, the late-time X-ray lightcurve is a power-law between $L_X\propto t^{-4/3}$ and $t^{-8/3}$.

The majority of the mass inflow is lost from the disk in the form of a radiatively driven wind.
The energy generated by accretion in the inner disk is expected to be reprocessed by this wind and released at the photon trapping radius, $r_{\rm{tr}}$, where the radiative diffusion time equals to the expansion time \citep[typically occurring a few to 10 days after the TDE;][]{Kremer2019c}. A rough estimate for the trapping radius is
\begin{multline}
    r_{\rm{tr}} \approx \min\left(v_{\rm w} t, {\dot{M}(t)\kappa \over 4\pi c}\right)\\
    \approx 10^{15}\mathrm{cm}\,\min\left({v_{\rm w}\over 10^9\rm\, cm\,s^{-1}} {t\over 10^6\rm\,s},\ {\dot{M}(t)\over 17M_\odot \, \mathrm{yr^{-1}}}\right),
\end{multline}
where $v_{\rm w}\approx 10^9\rm\, cm\, s^{-1}$ is the typical wind speed \citep{Kremer2019c} and $\kappa= 0.34\rm\, cm^2\, g^{-1}$ is the electron scattering opacity for solar composition material.

For a detailed discussion of the radiation hydrodynamics of the accretion disk and wind, we direct the reader to \citet{Kremer2019c,PiroLu2020} and references therein. Here, we summarize the key points. As a result of adiabatic loss, the emerging luminosity is smaller than the accretion luminosity by a factor of $(r_d/r_{\rm tr})^{2/3}\sim 10^{-2}$ for $r_{\rm tr}\sim 10^{15}\rm\, cm$ and $r_d\sim 10^{12}\rm\, cm$ roughly at $t\sim 10\,$days. \citet{Kremer2019c} considered $s\in (0.2, 0.8)$ and found the peak bolometric luminosity to be in the range $10^{41}-10^{44}\,\rm{erg\,s}^{-1}$ (depending also upon the assumed stellar parameters, such as masses) and the spectrum to be in the optical/UV, for typical stellar-mass BH TDEs. This optical emission is expected to last roughly $10-100$ days until the mass inflow rate drops below roughly $(r_d/r_g)L_{\rm{Edd}}/c^2$, at which point the disk is expected to transition to a geometrically thin state and the accretion rate may drop by many orders of magnitude \citep{ShenMatzner2014}. 

In Figure \ref{fig:TDE}, we summarize the key evolutionary features of the first roughly 100 days following the tidal disruption.

\subsection{Comparison to Fast Blue Optical Transients}\label{sec:compare_to_FBOTs}

\begin{deluxetable*}{l|c|l}
\tabletypesize{\small}
\tablewidth{0pt}
\tablecaption{Comparison of key features of stellar-mass BH TDEs and FBOTs}
\tablehead{
    \colhead{} &
    \colhead{ Stellar-mass BH TDEs} &
	\multicolumn{1}{c}{FBOTs}
}
\startdata
Peak optical luminosity $[\rm{erg\,s}^{-1}]$ & $\approx10^{41}-10^{44}$ & \textbf{Overall:} $\approx10^{41}-10^{44}$ \scriptsize{ \citep{Drout2014,Pursiainen2018} }\\
 & & \textit{AT2018cow:} $\approx 4\times10^{44}$ \scriptsize{ \citep{Margutti2019} } \\
 & & \textit{ZTF18abvkwla:} $\approx 10^{44}$ \scriptsize{\citep{Ho2019}} \\ 
\hline
Optical rise time [days]& $\approx\,\rm{a\,\,few}-10$ &  \textbf{Overall:} $\lesssim 5$ \scriptsize{\citep{Drout2014,Pursiainen2018} } \\
 & & \textit{AT2018cow:} $1.43 \pm 0.08$ \scriptsize{\citep{Prentice2018,Perley2019} } \\
 & & \textit{ZTF18abvkwla:} $1.83 \pm 0.05$ \scriptsize{\citep{Ho2019}} \\
 \hline
 Fade time [days]& $\approx\,$a few &  \textbf{Overall:} $\sim\,\rm{a\,few}-10$ \scriptsize{\citep{Drout2014,Pursiainen2018} } \\
 & & \textit{AT2018cow:} $1.95 \pm 0.06$ \scriptsize{\citep{Prentice2018,Perley2019} } \\
 & & \textit{ZTF18abvkwla:} $3.12 \pm 0.22$ \scriptsize{\citep{Ho2019}} \\
 \hline
X-ray luminosity $[\rm{erg\,s}^{-1}]$ & (assuming unabsorbed) & (Observed values are for $0.3-10\,$keV) \\ 
\qquad $\approx 1\,$d after peak & $\approx10^{43}-10^{47}$ & \textit{AT2018cow:} $\approx 10^{43}$ \scriptsize{\citep{RiveraSandoval2018} } \\
\hline
\qquad $\approx 10\,$d after peak & $\approx10^{41}-10^{46}$ & \textit{AT2018cow:} $\approx 5\times 10^{42}$ \scriptsize{\citep{RiveraSandoval2018} } \\
\hline
\qquad $\approx 100\,$d after peak & $\approx10^{38}-10^{45}$ & \textit{AT2018cow:} $\approx 10^{40}$ \scriptsize{\citep{Margutti2019} } \\
 & & \textit{CSS161010:} $\approx 5\times10^{39}$ \scriptsize{\citep{Coppejans2020} } \\
\hline
Volumetric rate [$\rm{Gpc}^{-3}\,\rm{yr}^{-1}$] & $20-200$ & $<560$ \scriptsize{ \citep[For $M_g < -20$;][]{Ho2020} }\\
 & & $700-1400$ \scriptsize{ \citep[For $M_g < -19$;][]{Coppejans2020} } \\
 & & $\gtrsim1000$ \scriptsize{ \citep[For $-15.8 < M_g < -22.2$;][]{Pursiainen2018} }\\ 
\enddata
\tablecomments{Summary of key features of stellar-mass BH TDEs alongside inferred FBOT properties from various references in the literature. For optical luminosity and rise time, we show ``Overall'' properties of the full FBOT population. For other features we list only observations from specific FBOTs, namely AT2018cow, CSS161010, and ZTF18abvkwla. The upper and lower bounds for theoretical X-ray luminosities of TDEs assume $s=0$ and $s=1$ power-law indices for the accretion rate, assuming no absorption; see Section \ref{sec:EMfeatures}. We show inferred CSM densities for a single epoch of radio observation for each observed FBOT.
\label{table:comparison}
}
\end{deluxetable*}

Recent high-cadence surveys have uncovered a growing number of fast-evolving transients with a wide range of observed properties. One class of particular interest is the Fast Blue Optical Transients \citep[FBOTs;][]{Drout2014}, also known as Fast Evolving Luminous Transients \citep[FELTs;][]{Rest2018}. Although a clear understanding of FBOTs remains elusive, this class of transients is generally defined by rise times (of order one to a few days) and peak luminosities (roughly $10^{41}-10^{44}\, \rm{erg\,s}^{-1}$) that are too fast and too luminous to be explained by the radioactive decay of $^{56}\rm{Ni}$. The majority of FBOTs are found in star-forming galaxies of roughly solar metallicity. Furthermore, the explosion sites span a range of off-sets from the galaxy centers, most closely resembling the off-set distribution of core-collapse supernovae \citep[e.g.,][]{Drout2014}.

The majority of FBOTs have been identified via archival searches of various optical surveys including the Pan-STARRS1 Medium Deep Survey \citep[PS1-MDS;][]{Drout2014}, the Dark Energy Survey \citep{Pursiainen2018}, Kepler \citep{Rest2018}, the Supernova Legacy Survey \citep{Arcavi2016}, and the Zwicky Transient Facility \citep[ZTF;][]{Ho2020}. In addition to the archival searches, a handful of FBOTs have been discovered while still active, notably AT2018cow \citep[also ATLAS18qqn;][]{Smartt2018,Prentice2018, RiveraSandoval2018,Margutti2019}, ZTF18abvkwla \citep{Ho2020}, and CSS161010 \citep{Coppejans2020}, enabling X-ray and/or radio follow-up.

Various analyses have computed volumetric rates of transients similar to FBOTs, with estimates ranging from roughly $100\,\rm{Gpc}^{-3}\,\rm{yr}^{-1}$ to more than $1000\,\rm{Gpc}^{-3}\,\rm{yr}^{-1}$ in the local universe \citep[e.g.,][]{Drout2014,Pursiainen2018,Ho2020,Coppejans2020}. Although the precise rate remains uncertain, the general consensus appears to be that FBOTs are roughly two to three orders of magnitude rarer than standard core-collapse supernovae \citep[e.g.,][]{Botticella2008}.

The specific origin of FBOTs remains unknown with a number of channels having been proposed including TDEs by IMBHs \citep{Perley2019}, massive star collapse and BH formation \citep{Quataert2019}, electron capture collapse following a white dwarf merger \citep{LyutikovToonen2019}, and magnetar formation \citep{Margutti2019}. Additionally, \citet{PiroLu2020} pointed out that many FBOT features (specifically in the case of AT2018cow) show similarities to what would be expected for a wind-reprocessed transient, regardless of the specific central engine.

Here, we propose stellar-mass BH TDEs as another possible FBOT progenitor. The rise times and peak optical luminosities predicted for these TDEs \citep[Section \ref{sec:EMfeatures} and][]{Kremer2019c} occupy the same region of parameter space expected for FBOTs, as does the estimated X-ray luminosity.\footnote{We do not consider here the variability in the X-rays observed on few day timescales in the case of AT2018cow \citep{RiveraSandoval2018}, but note that this variability may in principle arise through orbital decay during a TDE.} If these TDEs occur in high-metallicity young stellar clusters \citep[expected to be a dominant site for star formation;][]{LadaLada2003}, the host galaxy type and off-set distribution for observed FBOTs may also be recovered. Furthermore, the rate we predict for stellar-mass BH TDEs in YSCs (up to roughly $200\,\rm{Gpc}^{-1}\,\rm{yr}^{-1}$ in the local universe) is comparable to the observationally-inferred FBOT rates.
In light of these similarities, stellar-mass BH TDEs that occur in YSCs may \textit{in principle} be a viable progentor for FBOT-like transients. In Table \ref{table:comparison}, we summarize for comparison the key features of both FBOTs and stellar-mass BH TDEs.

Although it remains to be determined whether bright radio emission is a defining feature of the FBOT class broadly, radio emission (consistent with self-absorbed synchrotron radiation) is observed for the AT2018cow, ZTF18abvkwla, and CSS161010 events. This radio emission suggests the presence of circumstellar medium with densities ranging from roughly $10-10^6\,\rm{cm}^{-3}$, for various observation epochs and for a range of microphysics assumptions \citep{Margutti2019,Ho2020,Coppejans2020}. If prior to its tidal disruption, a main sequence star is tidally captured by a BH, such that a small amount of orbital energy is deposited into the star \citep[e.g.,][see left-most panel of Figure \ref{fig:TDE} for illustration]{Fabian1975}, a small amount of debris may be unbound from the star. This may occur either through partial stripping by the BH at the first pericenter passage or during subsequent pericenter passages if the star's envelope expands due to energy injected through the tidal encounter. The circumstellar medium density inferred from radio emission may be produced by a small amount of ejecta ($\lesssim10^{-2}M_\odot$) homologously expanding for a few years at a speed of $10^3\,\rm km\,s^{-1}$. More detailed hydrodynamic models of the disruption process are required to calculate the radial density profile and predict the radio emission.

\section{Conclusions \& Discussion}
\label{sec:conclusion}

\subsection{Summary}

We summarize here the main findings of this work:

\begin{enumerate}
    \item Using a suite of roughly $3000$ $N$-body simulations, we have explored the rates and properties of stellar-mass BH TDEs in YSCs. We derived TDE rates as a function of cluster mass and showed that the rates derived from the models agree closely with rates derived from simple analytic estimates.
    
    \item Using the rate scalings derived from our models and integrating over the full cluster mass function, we predict these TDEs occur at a rate of roughly $20-200\,\rm{Gpc}^{-3}\,\rm{yr}^{-1}$ in the local universe. The range in this estimate is determined primarily by the assumptions made concerning initial cluster densities \citep[e.g.,][]{PortegiesZwart2010,Marks2012}. Overall, we find TDEs occur a factor of a few to 10 times more frequently in binary-mediated dynamical encounters compared to single--single encounters.
    
    \item In YSCs of roughly solar metallicity, we predict a median BH mass of roughly $10\,M_{\odot}$. In lower-metallicity ($Z \lesssim 0.1Z_{\odot}$) clusters that may have formed at higher redshift, we predict a median mass of roughly $30\,M_{\odot}$. This difference is primarily a result of the metallicity-dependent stellar wind mass loss \citep{Vink2001}.
    
    \item We showed that stellar-mass BH TDEs exhibit the following key features: (i) rise times of roughly a day driven by the viscous accretion timescale, (ii) peak X-ray luminosities of roughly $10^{44}-10^{48}\,\rm{erg\,s}^{-1}$, and (iii) optical luminosities of up to roughly $10^{44}\,\rm{erg\,s}^{-1}$ produced by re-processing of X-rays by a disk wind on a timescale of a few to 10 days after the TDE. On the basis of all of these features, combined with the estimated TDE rates and host-galaxy properties, we propose stellar-mass BH TDEs as a viable progenitor for the FBOT class of transients \citep[e.g.,][]{Drout2014,Margutti2019,Ho2020,Coppejans2020}.
    
\end{enumerate}

\subsection{Discussion \& Future Work}

Stellar-mass BH TDEs distinguish themselves from other proposed channels for FBOTs (see Section \ref{sec:compare_to_FBOTs}) in the combination of X-ray and optical signatures. For instance, in the case of stellar collapse, long-lasting power from a central engine may be supplied by fallback accretion onto a BH or magnetar spindown, where the late-time energy injection rates are expected to be $L\propto t^{-5/3}$ and $L\propto t^{-2}$, respectively. In our model for stellar-mass BH TDEs, the energy injection rate may have a broad range of energy injection rates from $L\propto t^{-4/3}$ to $L\propto t^{-8/3}$. This leads to different late-time ($t\gtrsim 10\rm\, d$) behaviors in the lightcurve that can in principle to used to distinguish our model from others. Of course, the detection of a host YSC may also hint at the proposed TDE scenario \citep[for example see][for a discussion of host features of AT2018cow]{Lyman2020}. However, core-collapse supernovae will also occur in sufficiently young star clusters, thus the presence of a host cluster may not provide indisputable evidence of a TDE origin.

Although the event rates inferred from FBOT observations are uncertain and vary between different analyses, in general, the observed FBOT rate appears higher than the stellar-mass BH TDE rate estimated from our models by a small factor (see Table \ref{table:comparison}). However, there are several reasons why our estimated rate may underestimate the true TDE rate, possibly explaining this discrepancy:

First, in this study, we have assumed zero primordial binaries. BH binaries were allowed to form exclusively through three-body formation (see Section \ref{sec:methods}).
Stellar binaries are known to increase the rates of dynamical collisions and TDEs \citep[e.g.,][]{Fregeau2007}. In this case, our predicted TDE rates may be a lower limit. To test this, we ran 50 additional simulations (\texttt{o} in Table \ref{table:models}) that adopt a primordial binary fraction of $100\%$. Consistent with previous \texttt{CMC} studies \citep[e.g.,][]{Kremer2020}, secondary masses are drawn from a uniform distribution in mass ratio in the range $[0.1,1]$ and initial orbital periods are drawn from a log-uniform distribution \citep[e.g.,][]{Sana2012}. In total, these models yield a factor of roughly 5 more TDEs per cluster (all of which occur through binary-mediated encounters) compared to the models with comparable mass and zero primordial binaries (\texttt{f} in Table \ref{table:models}). Thus, primordial binaries may increase the TDE rates quoted in Table \ref{table:rates} by a small factor. We reserve for follow-up studies a more expansive investigation of primordial binaries and their implications for binary-mediated TDEs.

Second, as discussed briefly in Section \ref{sec:EMsignatures}, a fraction of TDEs may occur through tidal capture where the main sequence star may undergo multiple passages before ultimately being disrupted. As discussed in e.g., \citet{Fabian1975}, the cross section for tidal capture may be a factor of roughly a few times larger than that for tidal disruption. If the fate of the majority of tidal captures is a TDE, the total TDE rate may be higher than that estimated here by a factor of a few.

Of course, realistic YSCs exhibit a much larger range in properties than those considered here. For instance, in our $N$-body models, we have adopted a relatively narrow range in initial cluster sizes, which may in reality extend from roughly $0.1-10\,$pc or even more, depending on the cluster mass \citep[e.g.,][]{Krumholz2019}. Although our models were intended to capture the ``mean" of this distribution, this value is uncertain and, of course, lower density clusters would exhibit a lower TDE rate. For example, from the supplemental models \texttt{n} in Table \ref{table:models}, we find that the TDE rate for clusters with initial $r_v=2\,$pc is a factor of roughly 6 times lower than than the rates computed in the $r_v=1\,$pc models. If indeed $r_v=2\,$pc is more representative of the full distribution of YSCs, the volumetric rates shown in Table \ref{table:rates} may overestimate the true value.

Along similar lines, we have adopted a relatively narrow range in initial cluster masses that does not span the full distribution of observed YSCs. For example, YSCs in the Antennae are observed with masses of $10^6\,M_{\odot}$ \citep[e.g.,][]{PortegiesZwart2010} or more while observed clusters in massive elliptical galaxies can reach $10^7\,M_{\odot}$ \cite[e.g.,][]{Bastian2006}. Although of high potential interest \citep[see e.g.,][for a recent study of super-star clusters with \texttt{CMC}]{Rodriguez2020}, we reserve consideration of these high-mass systems for later study.

In Section \ref{sec:masses}, we showed that, depending on the assumed initial virial radius and cluster metallicity, IMBHs with masses in excess of $100\,M_{\odot}$ may form in YSCs through stellar collisions \citep[see also][]{DiCarlo2019,Kremer2020c} and ultimately undergo TDEs. The topic of TDEs by IMBHs in stellar clusters has been examined at length \citep[e.g.,][]{Rosswog2009,MacLeod2014,MacLeod2016,Fragione2020}. We have focused in this study on the more common (by a factor of roughly 10:1 or higher in our models) stellar-mass BH TDE. However, IMBH TDEs likely lead to several notable differences. For example, for sufficiently massive IMBHs, the TDEs will transition from viscosity driven to fallback driven, thus becoming qualitatively more similar to a supermassive BH TDE \citep[e.g.,][]{Rees1988}. We leave for future work a more careful examination of the properties and observational signatures of TDEs by IMBHs.

We have used here the \texttt{CMC} code to model the evolution of YSCs. The Monte Carlo-based approach used in \texttt{CMC} allows us to model large populations of clusters at low computational expense compared to direct $N$-body models \citep[e.g.,][]{Pattabiraman2013,Rodriguez2016b}. However, unlike massive globular clusters, the relatively low-mass YSCs studied here can also be studied efficiently with direct $N$-body models \citep[e.g.,][]{DiCarlo2019,Banerjee2017}. Future work may examine the topic of TDEs using direct $N$-body models which would allow detailed examination of several regimes Monte Carlo simulations like \texttt{CMC} are ill-equipped to study. For example, examination of the final stage of cluster's life as it dissolves on a dynamical timescale through its tidal boundary and examination of clusters that contain massive IMBHs that may affect the overall cluster and TDE dynamics. Additionally, direct $N$-body models are more suited to study the lowest mass stellar associations \citep[$N\lesssim 100$;][]{LadaLada2003}, for which the spherical symmetry assumptions at the heart of the Monte Carlo-based approach break down.

Finally, unlike supermassive BH TDEs which have been studied extensively with hydrodynamic simulations, the stellar-mass  BH regime has been little explored, with a few exceptions \citep{Perets2016,Lopez2018}. Ultimately, hydrodynamic models are necessary to understand the detailed features of these events. For example, in the analytic estimates in Section \ref{sec:EMsignatures}, we considered only those interactions near the tidal disruption boundary, $r_{\rm{TD}},$ of the star. However, given that the interaction cross section scales linearly with stellar radius (Equation \ref{eq:sigma_ss}), a fraction of such BH--star encounters may actually occur in the direct collision regime, where $r_p < R_{\star}$ \citep[e.g.,][]{FryerWoosley1998,HansenMurali1998}. Although this subclass of ``head on'' encounters is unlikely to affect our rate predictions by more than a small factor, these physical collisions may produce a subclass of unique transients. For instance, a direct collision may lead to prompt accretion (i.e., $t_{\rm{fb}}\approx 0$) onto the BH which, among other possible consequences, may make the effects of feedback critical on the subsequent evolution, if indeed a disk forms at all. On the other hand, as mentioned briefly in Section \ref{sec:EMsignatures}, even more distant encounters ($r_p > r_{\rm{TD}}$) may lead to tidal capture, possibly resulting in multiple passages that each partially strip the star \citep[e.g.,][]{Fabian1975,Ivanova2017}, again potentially producing EM signatures unique from those presented in the classic TDE regime discussed in Section \ref{sec:EMsignatures}.
More careful treatment of the various regimes of BH--star interactions with hydrodynamic models is necessary to explore the potentially broad range of outcomes of these events in greater detail.

\acknowledgements{We thank Tom Maccarone for helpful suggestions on the manuscript and Carl Rodriguez for useful preliminary discussions. We also thank the anonymous referee for careful review and many helpful suggestions. K.K. is supported by an NSF Astronomy and Astrophysics Postdoctoral Fellowship under award AST-2001751. W.L. is supported by the David and Ellen Lee Fellowship at Caltech. S.C. acknowledges support of the Department of Atomic Energy, Government of India, under project no. 12-R\&D-TFR-5.02-0200. F.A.R. and C.S.Y. acknowledge support from NSF Grant AST-1716762 at Northwestern University.}

\bibliographystyle{aasjournal}
\bibliography{mybib}

\end{document}